# Ta, Ti and Hf effects on Nb₃Sn high-field performance: temperature-dependent dopant occupancy and failure of Kramer extrapolation

Chiara Tarantini[1], Shreyas Balachandran[1], Steve M Heald[2], Peter J Lee[1], Nawaraj Paudel[1], Eun Sang Choi[1], William L. Starch[1] and David C Larbalestier[1]

[1] National High Magnetic Field Laboratory, Florida State University, Tallahassee, FL 32310, USA
[2] Advanced Photon Source, Argonne National Laboratory, Argonne, IL 60439, USA

E-mail: tarantini@asc.magnet.fsu.edu

## Abstract

The increasing demand for improving the high-field (16-22 T) performance of Nb₃Sn conductors requires a better understanding of the properties of modern wires much closer to irreversibility field, $H_{Irr}$. In this study we investigated the impact of Ta, Ti and Hf doping on the high-field pinning properties, the upper critical field, $H_{c2}$, and $H_{Irr}$. We found that the pinning force curves of commercial Ti and Ta doped wires at different temperatures do not scale and that the Kramer extrapolation, typically used by magnet designers to estimate high-field critical current density and magnet operational margins from lower field data, is not reliable and significantly overestimates the actual $H_{Irr}$. In contrast, new laboratory scale conductors made with Nb-Ta-Hf alloy have improved high-field $J_c$ performance and, despite contributions by both grain boundary and point defect pinning mechanisms, have more predictable high-field behavior. Using Extended X-ray Absorption Fine Structure spectroscopy, EXAFS, we found that for the commercial Ta and Ti doped conductors, the Ta site occupancy in the A15 structure gradually changes with the heat treatment temperature whereas Ti is always located on the Nb site with clear consequences for $H_{c2}$. This work reveals the still limited understanding of what determines $H_{c2}$, $H_{Irr}$ and the high-field $J_c$ performance of Nb₃Sn and the complexity of optimizing these conductors so that they can reach their full potential for high-field applications.

## 1. Introduction

Nb₃Sn provides the most technologically ready conductors for the realization of high field accelerator magnets beyond the limits of Nb-Ti. In fact, Nb₃Sn has been selected for the magnets employed in the Hi-Luminosity upgrade of the Large Hadron Collider (LHC)[1,2,3] and it is the first choice for the realization of the Future Circular Collider (FCC).[4] Because of the very demanding requirements of this project and broad interest in improving the performance for other applications like NMR spectroscopy, compact cyclotrons and magnetically confined fusion reactors, a better understanding of the limits of present commercially available conductor could provide insights for further improvements or development of new conductor designs. Because of the need to enhance the high field properties, investigation of pinning mechanisms and the variation of the irreversibility field, $H_{Irr}$, and the upper critical field, $H_{c2}$, are of great interest.

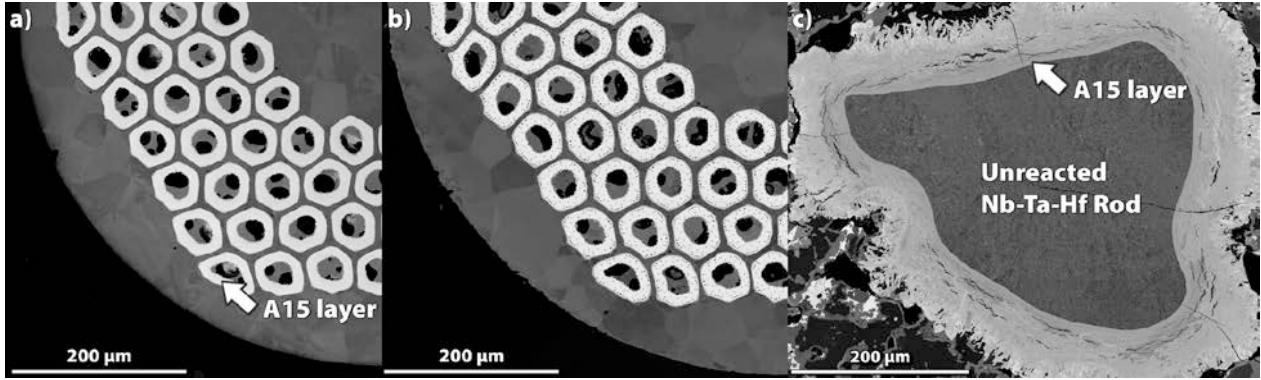

**Figure 1** FESEM backscattered electron images showing partial cross-sections of a) the Ta-doped RRP-13711 (48h/664 °C) wire, b) the Ti-doped RRP-149984 (48h/665 °C) wire and c) the monofilamentary Ta and Hf doped wire.

Historically the main vortex pinning mechanism considered responsible for Nb$_3$Sn properties is grain boundary (GB) pinning whose functional dependence is given by $F_{p,GB}(H) = A_{GB}\left(\frac{H}{H_{Irr}}\right)^{0.5}\left(1 - \frac{H}{H_{Irr}}\right)^2$.[5,6,7,8] Since the $H_{Irr}$ of Nb$_3$Sn conductors (~23-25 T) greatly exceeds the typical fields (~15 T) available in most laboratories, $H_{Irr}$ is typically estimated by extrapolation from mid-field data using the Kramer function $Kr(H) = J_c^{0.5}(\mu_0 H)^{0.25}$.[9] In fact the $Kr(H)$ function corresponds to the linearization of the $F_{p,GB}(H)$ function with $H_{Irr}$ defined as $Kr = 0$. Indeed a valuable correlation was found by the US Magnet Development Program between this $H_{Irr}$ extrapolation and the $I_c$(4.2 K, 15 T) performance of conductors, even though they assumed the same field-dependent $f_p(h)$ function to be valid at every temperature (with $f_p$ and $h$ being the reduced-$F_p$ and $H/H_{Irr}$) and employed it to estimate the operational margin.[10] However, the validity of the Kramer extrapolations as a means to predict $J_c(H,T)$ outside the 15 T range has not been experimentally verified at high fields in modern Ta/Ti-doped Nb$_3$Sn conductors where the dopants are invariably used to increase $H_{Irr}$ and $H_{c2}$.

Another area of interest is the recent discovery that the Nb$_3$Sn vortex pinning properties can be strongly modified at high fields by introducing group IVB dopants, such as Zr or Hf, together with Ta.[11] Zr or Hf are employed to enhance the pinning force density, $F_p$, whereas Ta is used to keep $H_{Irr}$ close to the values of commercial conductors. This approach led to an increase in the bulk pinning force maximum from values that are typically ~4.6 T to values over 5.8 T (at 4.2 K) and an improvement by a factor three in the layer critical current density at 4.2 K and 16 T for the Hf-doped wire without the need for internal oxidation by using SnO$_2$ (SnO$_2$ appears to be necessary in the Zr case). Avoidance of SnO$_2$ is particularly appealing for wire manufacturers because it would allow a more straightforward transition from lab-scale to commercial fabrication. Interestingly, pinning force measurements on the Hf and Zr wires indicate a strong contribution to pinning from a source other than grain boundaries, most likely point defects (PD), as inferred by the significant shift toward higher field of the $F_p$ maximum, despite small variation of $H_{Irr}$. The high irreversibility field of oxidized Ta-Zr doped laboratory-scale monofilaments was also confirmed by simultaneous work conducted on multifilamentary wires.[12] For future use of these newly alloyed wires, a better understanding of how to analyze and to predict their behavior would be valuable.

As mentioned above, Ta and/or Ti are the $H_{c2}$-enhancing dopants used in current Nb$_3$Sn production strand, and we may recall that, in bronze wires, Ta appeared to be only half as efficient as Ti maximizing $H_{c2}$, since it requires ~4 at.%Ta while less than 2 at.%Ti is needed.[13] For many years they both were assumed to occupy the Nb site in the A15 structure.[14,15,16] However, analyzing the overall A15 layer compositions and noticing the frequent Sn deficiency (<25%) in Ti-doped Nb$_3$Sn compared to the Ta case, it was suggested that Ta occupies the Nb site whereas Ti the Sn site.[17] To test this hypothesis, we employed EXAFS (Extended X-ray Absorption Fine Structure), which much to our surprise showed that Ti actually sits only on the Nb site, whereas Ta splits between the Nb and Sn sites.[18] This finding revealed that $H_{c2}$ is not simply determined by the dopant itself but also by the anti-site disorder (i.e. Nb on the Sn site and vice versa) that the dopant induces. Moreover, since a sensibly different overall composition and site occupancy were found in two different Ta-doped conductors (a 54/61 and a 108/127 restack designs with ~50 and 70 μm filament diameter, respectively) after different heat treatments, a possible effect of the heat treatment (HT) on site occupancy was hypothesized.

In this work we explore some of these ideas. By high-field characterization we investigate the temperature dependence of $F_p$ for a modern RRP® conductor to verify whether or not the Kramer extrapolation is a reliable way to determine $H_{Irr}$ and whether the GB functional dependence can be assumed at all temperatures. The same type of characterization was also employed to investigate how the pinning mechanisms change in Zr and Hf-alloyed conductors. We also suggest how to estimate $H_{Irr}$ when it is not possible to access high-field facilities. Finally we heat treated Ta and Ti-doped RRP®

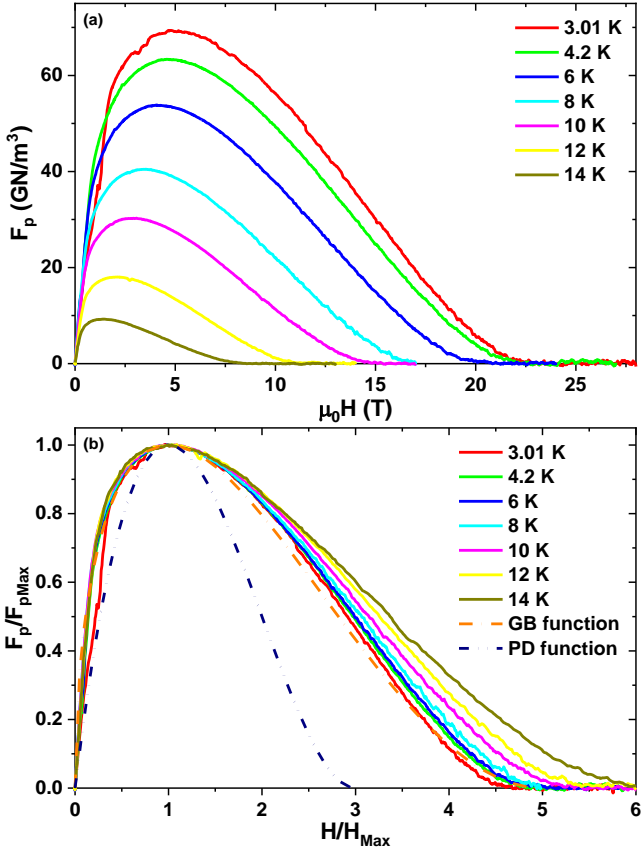

**Figure 2** (a) Field dependence of the non-Cu $F_p$ at different temperatures and (b) the same curves normalized to maximum for a Ta-doped RRP® wire. The dashed lines represent the theoretical function for grain boundary (GB) and point defect (PD) pinning mechanisms.

Nb$_3$Sn wires and used EXAFS to see whether or not the site occupancy changes with HT temperature. With these results and with specific heat characterization, the effects on $H_{c2}$ are also discussed.

## 2. Experimental

In this work we characterize both commercial multifilamentary strands and laboratory-scale monofilaments. First we studied two 0.82-0.85 mm diameter, 108/127 design large-scale production RRP® strands: a standard-Sn Ta-doped wire, heat treated at 634, 666 and 711 °C for 48 hours, and a reduced-Sn Ti-doped wire, heat treated for 48 hours at 5 different temperatures between 606 and 727 °C. The second sample type were monofilamentary wires made with three different alloys: Nb-4Ta, Nb-4Ta1Zr and Nb-4Ta1Hf by atomic percent. Each of these alloys was employed in wires made with two different Cu-Sn powder mixtures, one with and one without SnO$_2$, as described in ref. 11. These special alloyed wires underwent a 670 °C/100h A15 reaction. The best performing of these wires used a Ta-Hf alloy without SnO$_2$ and is the primary focus here. Figure 1 shows the Field Emission Scanning Electron Microscope (FESEM) images of the two RRP strands and of the Ta-Hf monofilamentary wire.

Magnetization hysteresis loops were performed in two vibrating sample magnetometers (VSM), one mounted in a 14 T superconducting magnet and one in the 35 T resistive magnet at the National High Magnetic Field Laboratory so as to enable the pinning force curves to be determined. To avoid the difficulties of determining *a priori* the irreversibility field $H_{Irr}$ to investigate the $F_p$ curves and to compare the shape with the expected theoretical trends, here we prefer to normalize the applied field $H$ to the position of the maximum, $H_{Max}$, obtaining the $F_p/F_{pMax}$ versus $H/H_{Max}$ plots (instead of the more commonly employed $F_p/F_{pMax}$ versus $H/H_{Irr}$). With this normalization the theoretical GB and PD curves go to zero at $H/H_{Max} = 5$ and 3, respectively (dashed lines in Figure 2b).[19]

EXAFS characterization occurs at x-ray energies above the absorption edge energy of the element under investigation. By tuning the x-ray energy to different absorption edges (like for Ta or Ti), the local structure surrounding specific elements in a complex material can be resolved. The measurements were made on beamline 20-ID at the Advanced Photon Source. More details can be found in ref. 18.

Specific heat characterizations were also performed in a 16 T Quantum Design physical property measurement system (PPMS) to determine $H_{c2}$ of selected wires as described in ref. 20 and 21.

## 3. Results

### 3.1 VSM characterization and analysis of commercial RRP® multifilamentary wire

The hysteresis loops of the Ta-doped RRP® wires were obtained in high magnetic fields. The $F_p$ curves of the sample heat treated at 666 °C were estimated in the 3-14 K range and

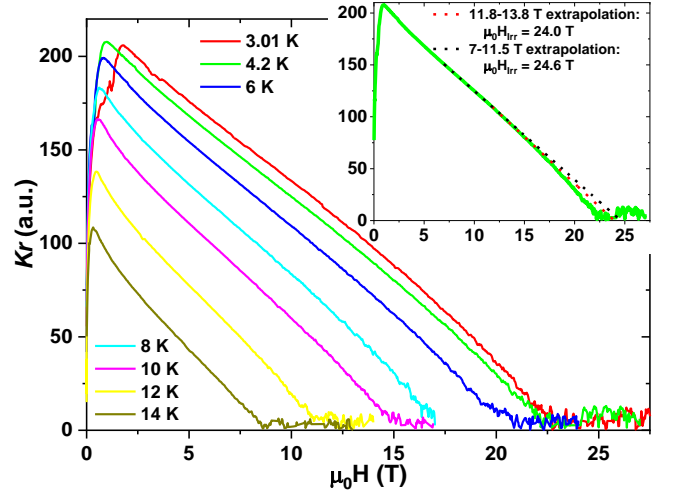

**Figure 3** Kramer plots at different temperatures for a Ta-doped RRP® wire. In the inset the 4.2 K data with extrapolations over different ranges (see main text).

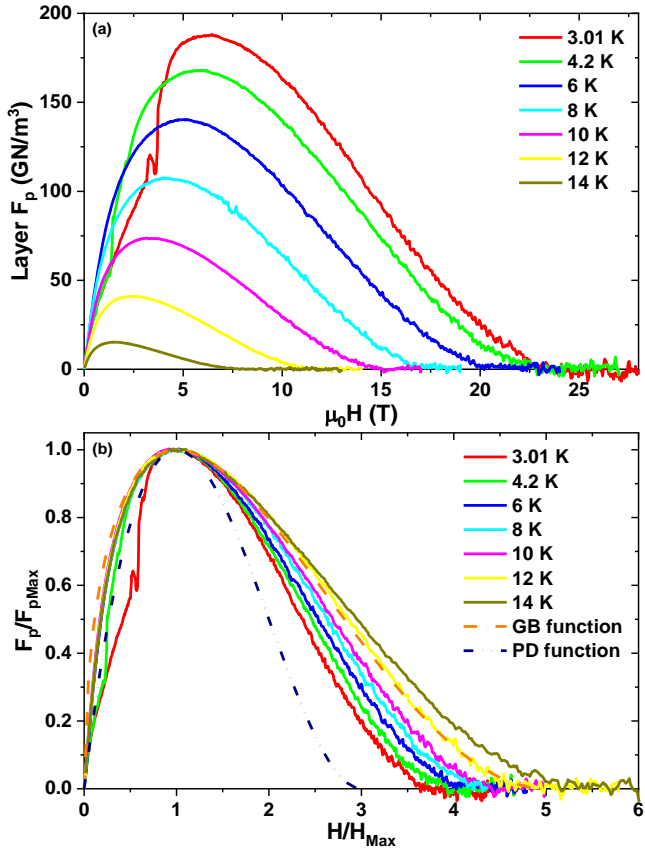

**Figure 4** (a) Field dependence of the layer $F_p$ at different temperatures and (b) the same curves normalized to the maximum for a Ta-Hf-doped monofilamentary wire (low-field, low-temperature data affected by flux jumps). The dashed lines represent the theoretical function for grain boundary (GB) and point defect (PD) pinning mechanisms.

are reported in Figure 2(a) as well as the same curves normalized to the peak position of $F_p$ in Figure 2(b). These plots reveal that the $F_p$ curves do not scale and that all high temperature data clearly exceed the GB function in the high field region. Only at 3 K do the data approach the GB function but there are still evident deviations.

As explained in the introduction, the typical method to analyze the superconducting properties is to use the Kramer plot $Kr(H) = J_c^{0.5}(\mu_0 H)^{0.25}$. Figure 3 reports these curves to verify the deviation from linearity of $Kr(H)$ and to estimate how much this affects the $H_{Irr}$ estimation. The main panel shows that the Kramer plots have marked upward curvatures at low fields (just above the maximum) and high temperatures, whereas at low temperature there are two opposite curvatures: upward in the low field region (below ~⅓$H_{Irr}$) and downward curvature in the high field region (from ~⅓$H_{Irr}$ to almost $H_{Irr}$). The field-range employed for the extrapolation to determine $H_{Irr}$ usually depends on the magnet available. A few examples are shown in the inset of Figure 3 for the 4.2 K curve. In our laboratory a 14 T VSM is routinely used for the wire characterization and, with a 11.8-13.8 T extrapolation range, a $\mu_0 H_{Irr}$ ~24 T would be estimated. Transport characterization frequently relies on smaller magnets[22] and a 7-11.5 T extrapolation leads to $\mu_0 H_{Irr}$ ~ 24.6 T. However, the present high-field characterization shows that the actual $\mu_0 H_{Irr}$ is about 22.4 T. This means that the mid-field extrapolations noticeably overestimate $H_{Irr}$ by more than 2 T (further analysis in given in section 4).

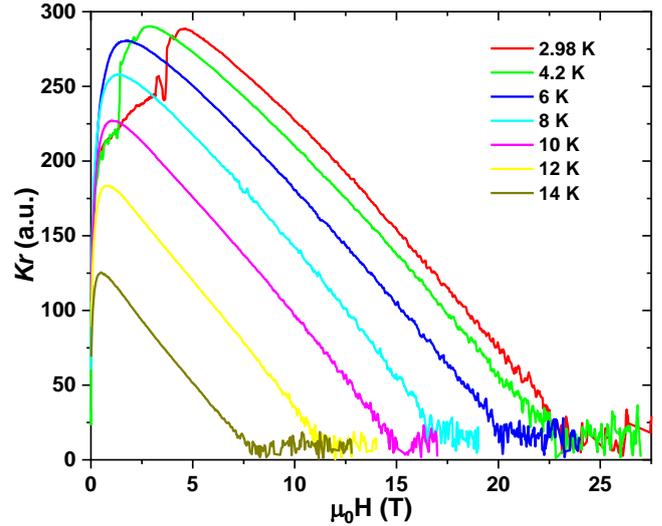

**Figure 5** Kramer plots at different temperatures for a high-$J_c$ Ta-Hf-doped monofilamentary wire.

Since we previously demonstrated that Ta-doped wires tend to have a larger distribution of properties than Ti-doped ones,[23] we also investigated by VSM a similarly heat-treated Ti-doped RRP® wire. The trend was overall the same of Figure 2(b) thus assuring us that these non-linear Kramer plot are not simply due to inhomogeneity of the $Nb_3Sn$ phase.

### 3.2 VSM characterization and analysis of special alloyed monofilamentary wire

Similar characterization was performed on the high-$J_c$ monofilamentary wires prepared with special alloys. Figure 4 shows the data for the wire manufactured with Nb-Ta-Hf alloy (without the use of $SnO_2$). It can be seen that also in this case there is a clear temperature dependence of the shape of the pinning force. However, differently from the RRP® samples, the normalized curves mostly lie between the GB and the PD functions suggesting a mixed contribution of those pinning mechanisms. In this case the Kramer plots (Figure 5) show a obvious downward curvature on substantially the entire field range. This means that, for these special alloyed wires, the Kramer plot cannot in any case be used to estimate $H_{Irr}$ because mid-field extrapolation would lead to large overestimations.

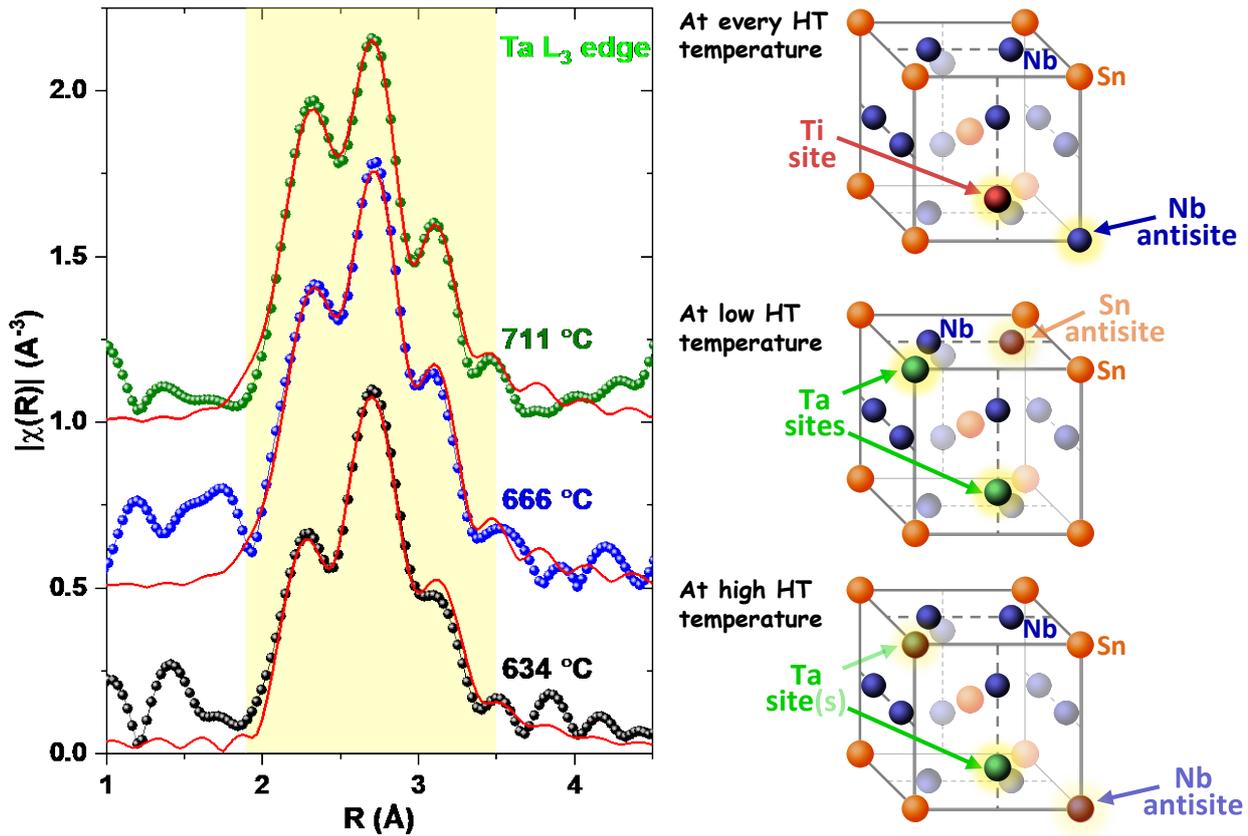

**Figure 6** (left) Fourier transforms of the $k^2$ weighted $\chi(k)$ data for the Ta $L_3$ edge using the $k$ range 2–11.5 for three differently heat treated Ta-doped RRP® samples. The points and line are the data, the red curves are the two-site fits (fitting window of interest for Nb$_3$Sn from 1.9 to 3.5 Å; the data are not phase corrected). The 666 and 711 °C curves are vertically shifted for clarity. (right) Crystalline structure sketches indicate that Ti always occupies the Nb sites, whereas Ta, which is also temperature dependent, splits between the Nb and Sn sites. The main type of antisite disorder that the dopants generate was demonstrated in ref. 18. The crystalline structure sketches are modified (by removing the indications of the emitting atoms and of the coordination shells and by adding site occupancies) from the Nb$_3$Sn structure schematic reported by Heald *et al.* in the center of Figure 1 in ref. 18 licensed under a Creative Commons Attribution 4.0 International License http://creativecommons.org/licenses/by/4.0/.

### 3.3 EXAFS characterization of RRP® wires

Nb and Sn sites in the A15 structure have different coordination shells.[18] The Nb site has three closely spaced nearest neighbor shells that produce a distinct 3-peak structure in the Fourier-transformed EXAFS spectrum. On the other hand the Sn site only has a single nearest neighbor shell generating a single peak in the Fourier-transformed EXAFS spectrum at the same position of the central peak of Nb three-peak structure.[18] This difference allows us to distinguish on which site the dopant sits. The experiment performed on the series of 5 Ti-doped samples reveal that Ti always sits on the Nb site independent of the heat treatment temperature, similarly to what we previously found on the single heat treatment studied in ref. 18. Those two studies do agree on there beings a strong preference of Ti for the Nb site. In contrast, a quite different behavior was found for the Ta-doped samples. Figure 6 shows the Fourier transformed data for the Ta L3 edge of the same Ta-doped conductor after different heat treatments. Despite all samples showing the three-peak structure suggesting that most of the Ta sits on the Nb site, it can be seen that the spectra are changing with the HT reaction. In particular the central peak, which is affected by Ta on both Nb and Sn sites, is decreasing in intensity with respect to the other two peaks that are produced by Ta occupying the Nb site. The EXAFS data in the main structure range (marked in yellow in Figure 6) were fitted using the Artemis software[24] with a model allowing the dopants to occupy either site. The best fits reveal a significant reduction of the Ta concentration on the Sn site with increasing heat treatment temperature. At 634 °C almost half of Ta, 43±7%, is on the Sn site. At 666 and 711 °C the Ta on the Sn site drops to 11±4 and 8±4%, respectively.

### 3.4 $H_{c2}$ characterization on RRP® wires by specific heat

Specific heat characterizations up to 16 T were performed on the Ta-doped RRP® wires in order to determine the temperature dependence of $H_{c2}$, as shown in Figure 7 together with the WHH[25] fits and the $H_{c2}$ slope near $T_c$. The estimated $H_{c2}(0)$ monotonically increases from about 26.7 to 28.1 T with increasing heat treatment temperature.

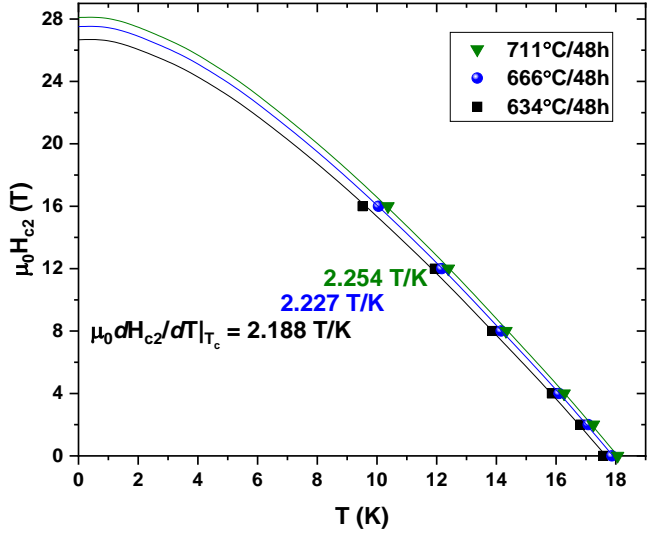

**Figure 7** Temperature dependence of $H_{c2}$ for the Ta-doped conductor after three different heat treatments. The lines correspond to the WHH fits for the three cases.

## 4. Analysis and Discussion

### 4.1 High-field characterization and $H_{Irr}$

Our high field results show that the typically used Kramer extrapolation fails to accurately predict $H_{Irr}$ for both standard ternary wires and in newly developed quaternary wires. To evaluate how poor the $H_{Irr}$ estimation is when using the Kramer extrapolations from data limited only to a low or mid-field range, we determined $H_{Irr}$ by linear fits over different field ranges. Figure 8 shows these $H_{Irr}$ estimations as a function of $\mu_0 H_{FitMax}$, the maximum fitting field, to simulate the lack of high field data. We used two fitting ranges for the extrapolation: a wide field range [from 4 T to $\mu_0 H_{FitMax}$ in (a) and from 2 T to $\mu_0 H_{FitMax}$ in (b), which substantially excludes only the low-field data near the Kramer plot maximum] and a 2 T-range below $\mu_0 H_{FitMax}$ [from ($\mu_0 H_{FitMax}$ - 2 T) to $\mu_0 H_{FitMax}$]. In both cases, because of the non-linearity, the Kramer extrapolations over a wide field range (full red squares) strongly overestimate (by several Tesla) the actual $H_{Irr}$ (represented by the grey shadow area), regardless of the fitting range. If we extrapolate using only a 2 T-range below $\mu_0 H_{FitMax}$ (open red squares) the error is in general reduced but, for the RRP wire, the error still remains ~0.5-2.5 T. A better estimation from a 2 T-range extrapolation can be obtained on the Ta-Hf-doped wire but only with data up to 18 T.

Other approaches can be attempted for a more reliable $H_{Irr}$ estimation from mid-field characterization. As explained earlier, the principal expected pinning mechanisms are by grain boundary (always present) and by point defects (in more recent pinning-engineered materials) whose respective $F_p$ field dependences are:

$$F_{p,GB}(H) = A_{GB} \left(\frac{H}{H_{Irr}}\right)^{0.5} \left(1 - \frac{H}{H_{Irr}}\right)^2 \quad (1)$$

$$F_{p,PD}(H) = A_{PD} \left(\frac{H}{H_{Irr}}\right) \left(1 - \frac{H}{H_{Irr}}\right)^2 \quad (2)$$

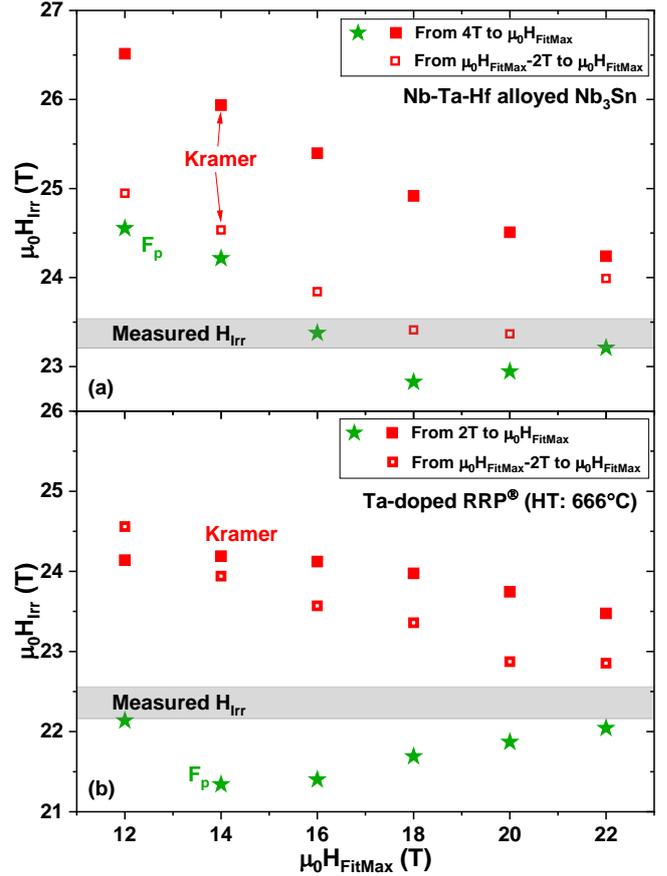

**Figure 8** $H_{Irr}$ as a function of the maximum fitting-field ($\mu_0 H_{FitMax}$) estimated from the standard Kramer function (squares) and from the $F_p$ curve fits (stars). Data are at 4.2 K for the Ta-Hf-doped monofilamentary wire (a) and for the Ta-doped RRP® wire heat treated at 666 °C (b). The horizontal shaded areas represent the actual $H_{Irr}$ value measured at high-field.

The generic equation

$$F_p(H) = A \left(\frac{H}{H_{Irr}}\right)^p \left(1 - \frac{H}{H_{Irr}}\right)^q \quad (3)$$

could in principle also be used to estimate $H_{Irr}$. However, fitting the $F_p(H)$ curve without high-field data can also generate large errors. In Figure 8 $F_p(H)$-fit estimations (green star symbols) are plotted as a function of $\mu_0 H_{FitMax}$. That both negative and positive errors exceeding 1 T occur depending on the fitting range make it clear that this approach is not accurate. It is also important to note that the lack of temperature scaling of the $F_p$ curves implies that the $p$ and $q$ parameters are actually temperature dependent and cannot be assumed constant.

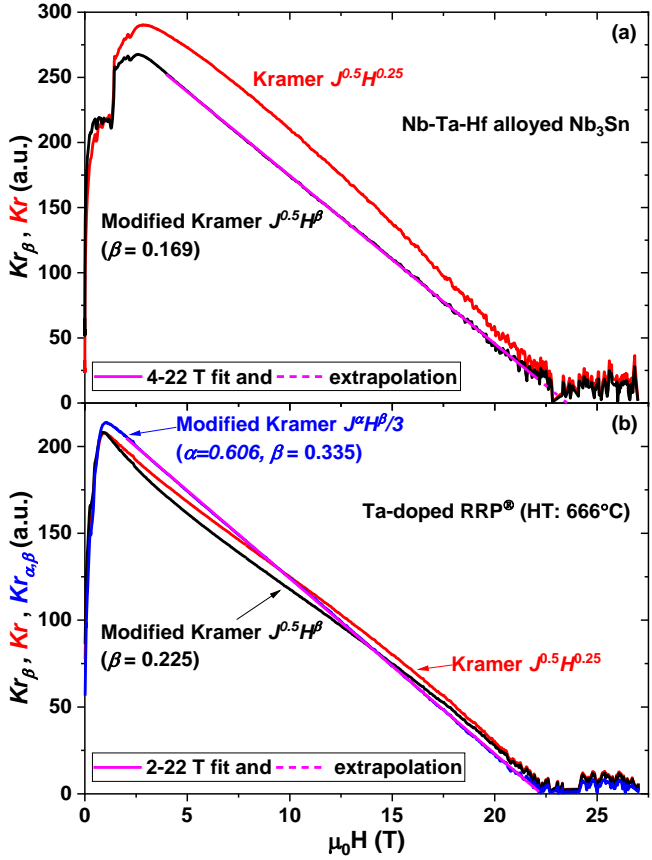
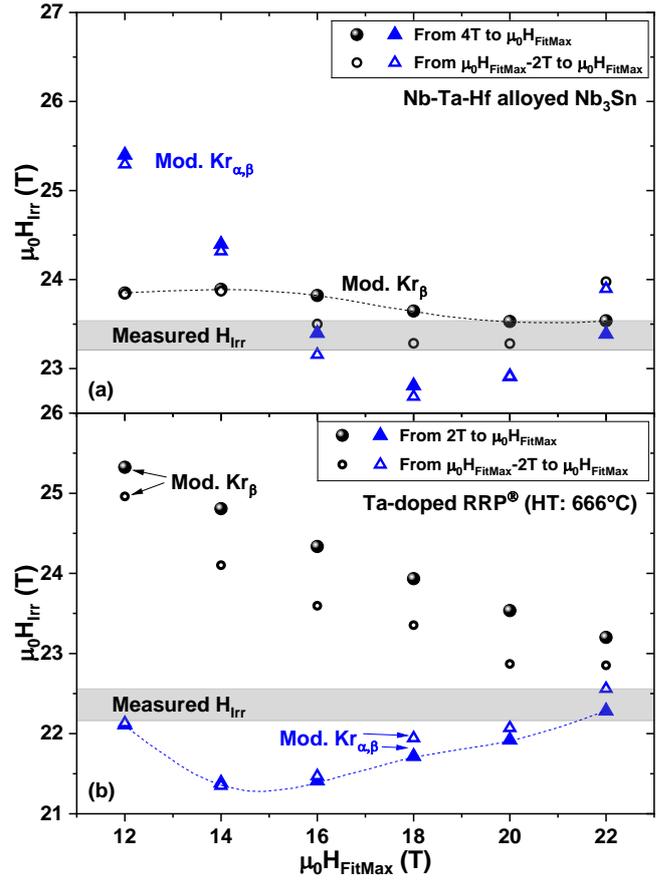

**Figure 9** Kramer function and modified Kramer functions at 4.2 K for the Ta-Hf-doped monofilamentary wire (a) and for the Ta-doped RRP® wire heat treated at 666 °C (b). The shown modified Kramer functions are the ones obtained by varying α and β (or only β) for the best linear fits in the 4-22 T and 2-22 T ranges (panels (a) and (b), respectively). See text for details.

**Figure 10** $H_{Irr}$ as a function of the maximum fitting-field ($\mu_0 H_{FitMax}$) estimated from the modified Kramer functions $Kr_\beta(H)$ (circles) and $Kr_{\alpha,\beta}(H)$ (triangles). Data are at 4.2 K for the Ta-Hf-doped monofilamentary wire (a) and for the Ta-doped RRP® wire heat treated at 666 °C (b). The horizontal shaded areas represent the actual $H_{Irr}$ value estimated at high-field. See text for details.

The linearized GB function [eq.(1)] is the Kramer function $Kr(H) = J_c^{0.5}(\mu_0 H)^{0.25}$ but a more generalized relation such as $Kr_{\alpha,\beta}(H) = J_c^\alpha (\mu_0 H)^\beta$ can be considered. Varying $\alpha$ and $\beta$ is equivalent to changing $p$ and $q$ in eq.(3), since $p$ and $q$ are related to $\alpha$ and $\beta$ by the relation $\alpha = 1/q$ and $\beta = (1-p)/q$. However, a simplified case may be usefully considered. In fact a linearization function for PD pinning [eq.(2)] assumes the form $Kr_{PD}(H) = J_c^{0.5}(\mu_0 H)^0 = J_c^{0.5}$. So in the case of mixed GB and PD pinning mechanisms, typical of some modern artificial-pin wires, a phenomenologically modified Kramer function could be expressed as $Kr_\beta(H) = J_c^{0.5}(\mu_0 H)^\beta$, explicitly varying $\beta$ to obtain a linear trend. Those mixed GB and PD cases are easily recognizable by a strong downward curvature of the Kramer plot over the entire field range (leading to the strong $H_{Irr}$ overestimation shown in Figure 8(a)), like the Ta-Hf case.

In Figure 9(a) we apply the $Kr_\beta(H)$ expression to the Ta-Hf sample and compare it to the Kramer function. $Kr_\beta(H)$ generates more consistent $H_{Irr}$ values, within 0.5 T accuracy even when the fitting range is limited to low fields [black circles in Figure 10(a)]. Figure 9(a) shows the good linear trend obtained over a wide field range (4-22 T), indicating that this method can be usefully applied to this type of wire with engineered pinning centers even when mid-field data are available. We attempt the same approach also for the RRP® wire [black curve and dots in Figure 9 (b) and 10(b)] however, varying only $\beta$ emphasizes the low-field upward curvature mentioned before with the result that the estimated $H_{Irr}$ is only marginally better than the Kramer extrapolation when extending the fit above 18 T. In both samples, determining the best $\beta$ over a wide fitting-range but extrapolating $H_{Irr}$ only from the 2 T-range below $\mu_0 H_{FitMax}$ (open black circles in Figure 10) results in a better estimation but, for the RRP sample, still not better than the similarly obtained Kramer extrapolation.

Although the most generic $Kr_{\alpha,\beta}(H)$ expression, varying both $\alpha$ and $\beta$, can be considered, this does not produce more accurate $H_{Irr}$ estimations for the Ta-Hf sample [blue triangles in Figure 10(a) are no better than the $F_p$ estimated ones in Figure 8] and only marginally better values are found for the RRP sample [blue curve and triangles in Figure 9(b) and 10(b)].

In summary, the results in Figure 8, 9 and 10 show that a reliable estimation of the true $H_{Irr}$ requires using fields well above those typically available in most laboratories. Most importantly, the commonly used Kramer extrapolation from mid-field cannot be used to determine the irreversibility field, since it strongly overestimates the actual $H_{Irr}$ values by 2-3 Tesla. The Kramer extrapolation should thus be considered accurate only in those rare cases where a linear trend can be observed over a wide temperature- and field-range (from just above the Kramer plot maximum up to the maximum available field). In the case of mixed GB/PD pinning a modified Kramer function of the type $Kr_\beta(H) = J_c^{0.5}(\mu_0 H)^\beta$ allows a more accurate $H_{Irr}$ estimation, within an 0.5 T error, even with data limited to 12 T; a correct estimation can be obtained with data up to 16 T (from the 2 T-range extrapolation). For more conventional wires, like RRP, $Kr_\beta(H)$ does not work well and the more general $Kr_{\alpha,\beta}(H) = J_c^\alpha(\mu_0 H)^\beta$ expression provides estimates similar to fitting the $F_p$ curve with 4 free parameters with significant variation on the $H_{Irr}$ estimation depending on the maximum field available. This means that in conventional conductors there is no reliable way to determine the true $H_{Irr}$ from mid-field data and high-field characterization is required.

$H_{Irr}$(4.2 K) was also evaluated at high-fields on the other two Ta-doped RRP samples, revealing a clearly reduced $H_{Irr}$ after the 634 °C HT (~20.3 T) and a mildly suppressed value after the 711 °C HT (~22.0 T) when compared to the ~22.4 T obtained with the 666 °C HT. This implies that the Nb-Ta-Hf alloyed Nb$_3$Sn does not have a suppressed actual $H_{Irr}$ as was observed in Nb-Zr alloyed Nb$_3$Sn wire.[26,27] In fact we measured $\mu_0 H_{Irr}$(4.2 K) ~23.4 T, about 1 T larger than the best Ta-doped RRP sample studied in this work, making Nb-Ta-Hf alloy a promising material for high-field Nb$_3$Sn optimization.

### 4.2 EXAFS and $H_{c2}$ characterizations

The change of dopant site occupancy by varying heat treatment temperature is a new observation that adds further complexity to the issue of how dopants disorder the A15 structure and raise $H_{c2}$. In our previous work[18] we postulated that the higher $H_{c2}$ of Ti-doped Nb$_3$Sn with respect to the Ta-doped Nb$_3$Sn is that Ti, sitting only on the Nb site, induces more anti-site disorder than the site-splitting Ta. The heat treatment study performed here on the same Ta-doped conductor indicates that Ta site occupancy is variable. Increasing the HT temperature causes the amount of Ta on the Sn site to drop from 43% at 634 °C to 11% at 666 °C and then to only 8% at 711 °C. Thus, Ta behaves at higher temperature more like the Ti dopant in increasing the disorder. Such disorder increase was verified by measuring the temperature dependence of $H_{c2}$ and estimating d($\mu_0 H_{c2}$)/dT at $T_c$, which is proportional to the normal state resistivity and so to the disorder. We found that d($\mu_0 H_{c2}$)/dT|$_{Tc}$ in fact increases from ~2.19, to ~2.23 and then to ~2.25 T/K, confirming the increasing disorder associated with sitting Ta on the Nb site (a small increase in $T_c$ was also found). Thus increasing the reaction HT temperature for Ta-doped strands not only improves the chemical homogeneity[20] but also provides the benefits of additional site disorder.

## 5. Conclusions

This work shows that in Nb$_3$Sn, despite being one of the most classical superconductors, new measurement techniques and alloys are providing us with a new understanding and suggest that there is the potential for further optimization.

We have previously shown by EXAFS that Ti dopant atoms only occupy the Nb sites in Nb$_3$Sn whereas Ta splits between Nb and Sn sites; in this study we have shown for the first time that the heat treatment temperature (in the 606-727 °C range) changes the dopant occupancy in the Nb$_3$Sn structure for Ta but not for Ti. The fraction of Ta on the Sn site drastically decreases when the heat treatment temperature increases from 634 to 711 °C. The increasing fraction of Ta atoms on the Sn site introduces more disorder into the structure and can be correlated to an increase in the $H_{c2}$ slope at $T_c$ and, as consequence, a larger $H_{c2}$ at low temperature.

The high field measurements reported here reveal that the pinning force curves taken at different temperatures do not scale ($p$ and $q$ being temperature dependent) and that the Kramer extrapolation is unreliable in all cases investigated, both for conventional large-scale production RRP conductor (both for Ta and for Ti doping) and laboratory-scale Ta-Hf alloyed monofilaments. In the Hf case however, due to the presence of mixed GB and PD pinning mechanisms, a reasonable estimation of $H_{Irr}$ can be obtained with a modified Kramer function of the type $Kr_\beta(H) = J_c^{0.5}(\mu_0 H)^\beta$ even with data limited to 12 T. In the case of RRP wires, no extrapolation can reliably estimate $H_{Irr}$ and high field characterization over the entire range is necessary. This implies that extensive characterization will be needed for the selection and the heat treatment optimization of conductors for the next generation of higher magnetic field designs. The high-field characterization also revealed that Ta-Hf doped Nb$_3$Sn presents both a shift in the pinning force maximum and an enhancement of $H_{Irr}$ above the Ta-doped Nb$_3$Sn values, making this materials of great interest for further development.

**Acknowledgements**

This work is funded by the U.S. Department of Energy, Office of Science, and Office of High Energy Physics under Award Number DE-SC0012083 and by CERN, and performed under the purview of the US-Magnet Development Program. This work was performed at the National High Magnetic Field Laboratory, which is supported by National Science Foundation Cooperative Agreements NSF DMR-1644779 and by the State of Florida. This research used resources of the Advanced Photon Source, an Office of Science User Facility operated for the U.S. Department of Energy (DOE) Office of Science by Argonne National Laboratory, and was supported by the U.S. DOE under Contract No. DE-AC02-06CH11357, and the Canadian Light Source and its funding partners.